# Impact of particle size on the magnetic properties of highly crystalline $Yb^{3+}$ substituted Ni-Zn nanoferrites

N. Jahan[1], M.M. Uddin[1*], M.N.I. Khan[2], F.-U.-Z. Chowdhury[1], M. R. Hasan[2], H. N. Das[2] and M.M. Hossain[1]

[1]Department of Physics, Chittagong University of Engineering and Technology (CUET), Chattogram 4349, Bangladesh
[2]Materials Science Division, Atomic Energy Center, Dhaka 1000, Bangladesh.

## ABSTRACT

Yb-substituted $Ni_{0.5}Zn_{0.5}Yb_xFe_{2-x}O_4$ ($0 \leq x \leq 0.20$ in the step of 0.04) ferrites have been synthesized using sol-gel auto combustion method. The structural characterization of the compositions has been performed by X-ray diffraction (XRD) analysis, field emission scanning electron microscopy (FESEM), quantum design physical properties measurement system (PPMS). That ensured the formation of single phase cubic spinel structure. Crystallite and average grain size are calculated and found to decrease with increasing $Yb^{3+}$ contents. Saturation magnetization ($M_s$) and Bohr magnetic moment ($\mu_B$) decrease while the coercivity increases with the increase in $Yb^{3+}$ contents successfully explained by the Neel's collinear two sub-lattice model and critical size effect, respectively. Critical particle size has been estimated at 6.4 nm from the $D_{XRD}$ vs. $M_s$, $H_c$ plot, the transition point between single domain regime (below the critical size) and multi-domain regime (beyond the critical size). Curie temperature ($T_c$) reduces due to the weakening of A-O-B super exchange interaction and redistribution of cations, confirmed by the M-T graph. The compositions retain ferromagnetic ordered structured below $T_c$ and above $T_c$, it becomes paramagnetic, making them plausible candidates for high temperature magnetic device applications. The relative quality factor (RQF) peak is obtained at a very high frequency ($\geq 10^8$ MHz), indicating the compositions could also be applicable for high frequency magnetic device applications.

Corresponding author: mohi@cuet.ac.bd

## 1. Introduction

Now-a-days, nanoferrites are very favorable materials due to their properties comparing with respected bulk ferrites, such as high surface to volume ratio, quantum confinement, super-paramagnetism, easy split-up under applying a magnetic field, and sturdy adsorption capacity. [1]. Spinel nanoferrites are very prominent materials due to their excellent properties that is used in technological and industrial applications such as biomedical, pharmaceutical, drug delivery, solar cells, gas sensors, magnetic information storage, MRI, Radar absorbing materials, supercapacitors, spintronics and microwave devices [2-4]. The size, shape, ionic nature, magnetic moments on the two sites (A- and B-sites) and electron hopping of the spinel nanoferrites significantly influence the magnetic properties as well as electrical properties of the materials [5]. Spinel nanoferrites having formula of $MFe_2O_4$ (M=$Ni^{2+}$, $Zn^{2+}$, $Mg^{2+}$, $Cd^{2+}$, $Mn^{2+}$, $Co^{2+}$ etc.) originate on the FCC structure comprises of 8 unit cell. Divalent ions occupy 8 sites among 64 tetrahedral A-sites, whereas 16 sites among 32 octahedral B-sites are occupied by trivalent ions [6]. Generally, three types of anti-ferromagnetic super exchange interaction such as $J_{AA}$ (A–O–A), $J_{AB}$ (A–O–B), and $J_{BB}$ (B–O–B) that occurs between cations on the A- and B-sites and mediated by oxygen ions, are mostly ruled the magnetic characteristics of spinel ferrites. The $J_{AA}$ contributes negligibly due to large separation between two *A*-site ions however, the $J_{AB} >> J_{BB}$. Various parameters can alter spin interaction: the difference in the size of cations, modification of cations distribution that changes the lattice constant and the oxygen parameter, thereby determining the $J_{AB}$.

The Ni-Zn nanoferrite is a versatile inverse spinel soft magnetic ferrite having the general chemical formula $(Zn^{2+}Fe^{3+})[Ni^{2+}Fe^{3+}]O_4$, where the Ni ferrite is considered as inverse spinel and Zn ferrite consider as normal spinel ferrite. $Zn^{2+}$ ions are non-magnetic and occupy the tetrahedral A-site and $Ni^{2+}$ prefer octahedral *B*-site but $Fe^{3+}$ occupy both A- and B-sites. The Ni-Zn possesses high magnetic permeability, high Curie temperature, high saturation magnetization and the values of low coercivity show~~s~~ high resistivity, negligible dielectric loss, low cost and chemically stable. Due to these versatile properties, the Ni-Zn ferrites are suitable for industrial and technological applications [7]. The electrical and magnetic properties can be enhanced by substituting the rare earth ions into the Ni-Zn ferrites. The physical properties and magnetic properties are drastically affected due to the larger ionic radius of the substituent rare earth (RE) ions have been reported by Rezlescu *et al.* [8]. They reported that the RE ions enter~~s~~ into the octahedral B-sites replacing the $Fe^{3+}$ ions, results in the exchange of 3d-4f coupling appears and changes the magnetization and Curie temperature. Moreover, the magnetic properties of Ni-Zn ferrite can be enhanced by modification of the magnetocrystalline anisotropy constant, A-O-B super exchange interaction, Yafet-Kittel (Y-K) angle, defect structure and microstructural evolution as well [9-16].

The magnetic properties of the RE doped nanoferrites strongly depend on the synthesis method, types of the substituent, cation distribution and the particle size [17]. There are different methods to synthesize the spinel soft nanoferrites such as sol-gel auto combustion, co-precipitation, hydrothermal, mechanochemical, microemulsion, thermal plasma synthesis, reverse micelle, sonochemical and rheological phase reaction [18]. Among them, the sol-gel auto combustion is preferable due to its high degree of homogeneity, low temperature, low cost and less time consuming [19]. As far our knowledge, no report has yet been published on the magnetic properties of RE ions $Yb^{3+}$ substituted Ni-Zn nanoferrites synthesized by sol-gel auto combustion method. In this report, the influence of particle size on the magnetic properties along with cation distribution, Curie temperature, magnetic properties mapping at elevated temperature, Y-K angles and P-E loop of sol-gel auto combustion technique mediated $Ni_{0.5}Zn_{0.5}Yb_xFe_{2-x}O_4$ ($x$= 0.00, 0.04, 0.08, 0.12, 0.16 and 0.20) nanoferrites has been presented..

## 2. Experimental procedure

Crystalline Yb-substituted Ni-Zn nanoferrites were synthesized by using sol-gel auto combustion technique. Analytical grade $Ni(NO_3)_2.6H_2O$(%), $Zn(NO_3)_2.6H_2O$(%), $Fe(NO_3)_3.9H_2O$(%) and $Yb(NO_3)_3.5H_2O$(%) (Merck Germany) were taken as precursor salts according to stoichiometric ratio for preparing the samples. The salts were dissolved into ethanol and stirred until it becomes homogeneous solution. The $p^H$ was fixed at 7 by mixing ammonia solution. Then the solution was kept on a magnetic heater at temperature 80°C to achieve dry gel. The dry gel was burnt into an oven at 250°C for 5 hrs. The dry ash was milled using a mortar pestle and achieved desired shape (ring and pellet) using hydraulic press (15 kN). Finally, the samples were sintered at 700°C for 5 hrs. Structural parameters were calculated from the XRD pattern which was carried out by using Rigaku Smart Lab with Cu-$K_\alpha$ (λ=1.5406 Å) radiation at room temperature (RT) in the range of 20 to 70°C and the scanning rate was 1° $min^{-1}$. Permeability measurements were performed using a Wayne Kerr precision impedance analyzer (6500B) in the frequency range of 10–120 MHz at RT. The driving voltage was set at 0.5 V. Using a quantum design physical properties measurement system (PPMS), magnetic properties were recorded at RT. P-E hysteresis curve was carried out by precision multiferroic test system (model: P-PMF, Radiant Tech. Inc. USA).

## 3. Results and Discussion

### 3.1 Structural analysis

The X-ray diffraction pattern of the series $Ni_{0.5}Zn_{0.5}Yb_xFe_{2-x}O_4$ ($x$=0.00, 0.04, 0.08, 0.12, 0.16 and 0.20) is shown in Fig. 1. The sharp and well defined peaks confirm the establishment of single phase spinel structure for all samples. There are no secondary peaks ~~are~~ observed in the pattern. Crystallite size, average grain size, lattice parameters and porosity have been calculated from the XRD spectra using established formalism and are presented in Table 1 that can also be found elsewhere [19]. The structural parameters such as theoretical cation distribution of tetrahedral and octahedral sites, ionic radius ($r_A$ and $r_B$), bond length ($R_A$ and $R_B$), tetrahedral, shared and unshared octahedral edge length (R, R′ and R″), oxygen position parameter (u) have also been calculated [19].

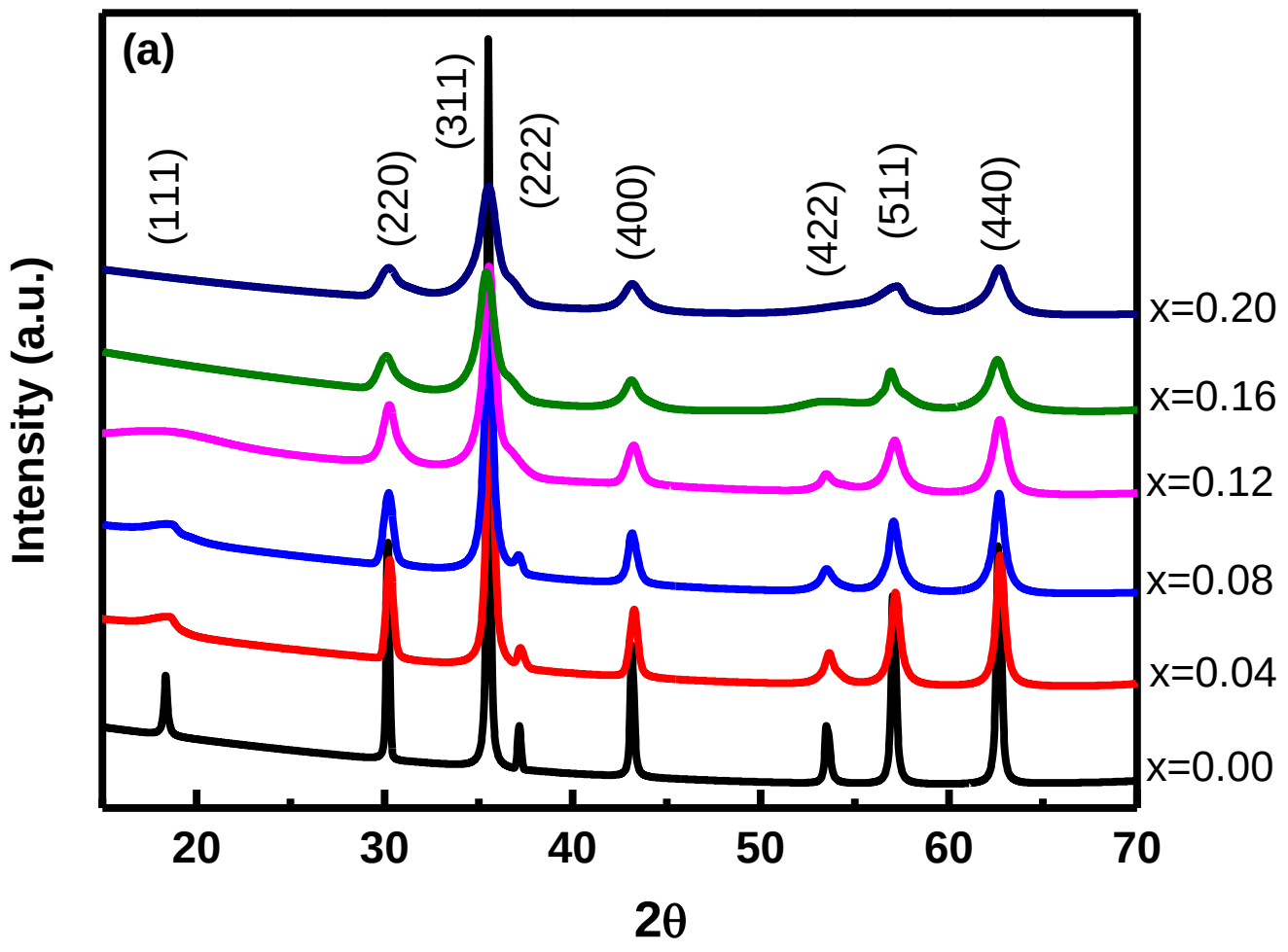

Fig. 1: The XRD pattern of $Ni_{0.5}Zn_{0.5}Yb_xFe_{2-x}O_4$ ($0\leq x \leq 0.20$ in the step of 0.04) compositions sintered at 700°C [19].

## 3.2 Microstructure study

The microstructure study is very important because the physical properties of the samples are strongly dependent on the grain size. The microstructure can be significantly tuned by changing the chemical composition. The structural studies have also made by a high resolution field emission scanning electron microscopy (FESEM) (JEOL JSM-7600F) as shown in Fig. 2. The average grain size of all the samples is measured using ImageJ software and is tabulated in Table 1. It is clear that as the substitution concentration $Yb^{3+}$ increases the grain size decreases as expected and more details in ref. [19].

Table 1: Values of crystallite size, average grain size, lattice parameter and porosity of $Ni_{0.5}Zn_{0.5}Yb_xFe_{2-x}O_4$ ($0 \leq x \leq 0.20$ in step of 0.04) compositions sintered at 700°C [19].

| Yb contents(x) | Crystallite size, (nm) | Average grain size , (nm) | Lattice parameter ($a_{expt}$) (Å) | Porosity P (%) |
|---|---|---|---|---|
| 0.00 | 23.1 | 52.06 | 8.393 | 55.15 |
| 0.04 | 10.9 | 24.29 | 8.388 | 66.60 |
| 0.08 | 9.5 | 24.12 | 8.395 | 69.11 |
| 0.12 | 6.4 | 18.97 | 8.398 | 70.18 |
| 0.16 | 4.8 | 17.97 | 8.397 | 71.34 |
| 0.20 | 5.6 | 17.50 | 8.396 | 72.48 |

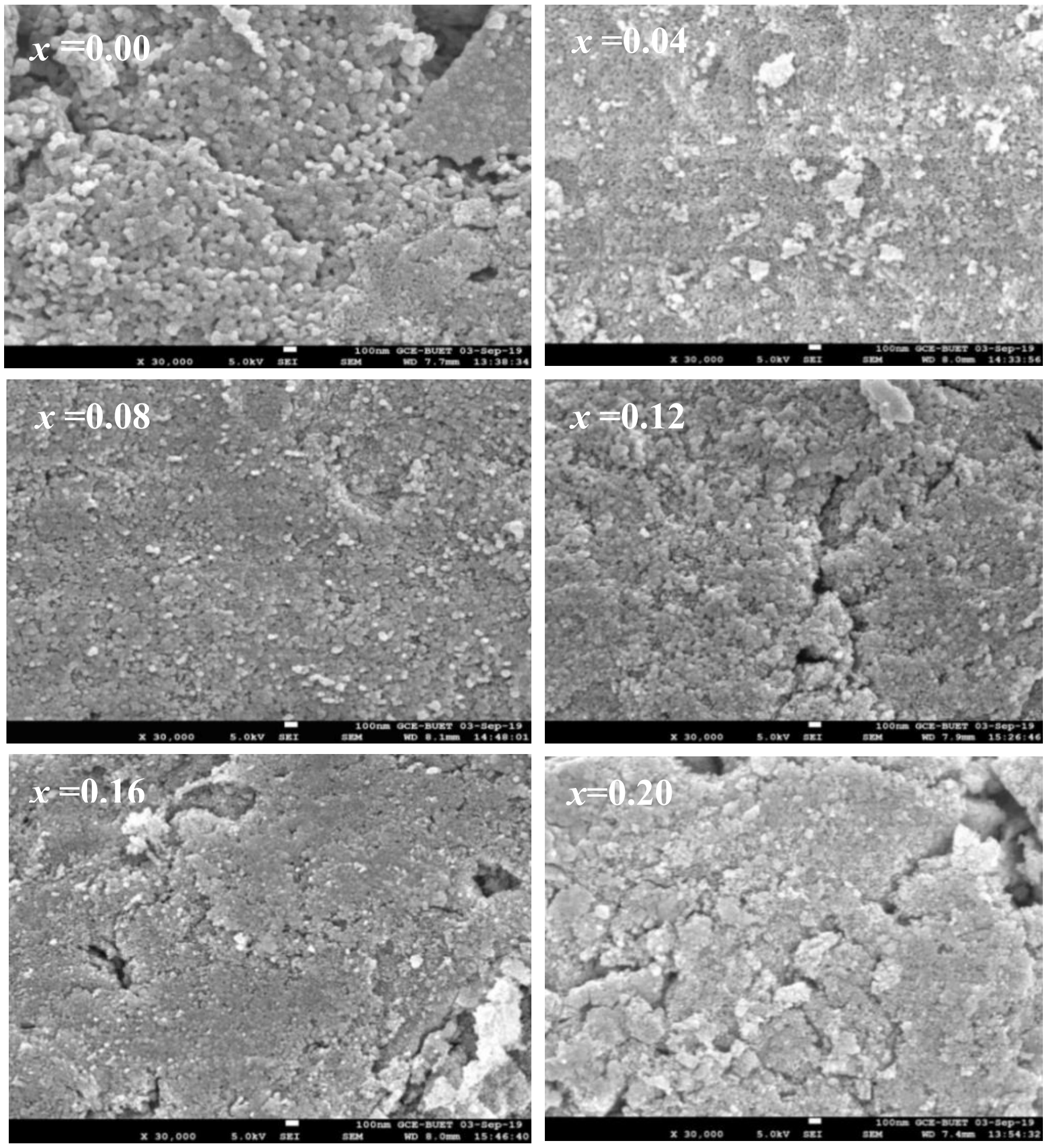

Fig. 2: The FESEM micrographs of $Ni_{0.5}Zn_{0.5}Yb_xFe_{2-x}O_4$ ($0 \leq x \leq 0.20$ in step of 0.04) compositions sintered at 700°C [19].

### 3.3 Magnetic properties

The shape and width of the hysteresis are affected by the physical parameters of the samples such as chemical composition of the compound, porosity, grain size etc. The technical important factor, ferrimagnetic nature of all samples can be established from the study of field dependent magnetization curve. The variation of magnetization (M) with the applied magnetic field (H), i.e., M-H curve has been measured at room temperature using the PPMS and illustrated in Fig. 3(a). All the samples demonstrate standard hysteresis pattern. It is known that the grain size, porosity and exchange interaction due to cation distribution have a significant influence on the shape and size of the hysteresis loop. It is renowned that at the beginning of applied field, the magnetic domain reoriented results the magnetization increases rapidly. After that due to the spin rotation, the magnetization gets slow and finally saturated. Magnetic parameters such as saturation magnetization ($M_s$), remanent magnetization ($M_r$), coercivity ($H_c$), squareness ratio ($SQR=M_r/M_s$) and anisotropy constant (K) have also been calculated using the hysteresis loop.

Saturation magnetization ($M_s$) of the samples has been evaluated using the extrapolation of magnetization (M) versus inverse of applied magnetic field (1/H) (inset of Fig 3a) and composition dependence of $M_s$ has also been illustrated in Fig. 3(b) [20]. The value of $M_s$ decreases (up to *x*=0.16) as the substituent atom ($Yb^{3+}$) increases which can be attributed by the cation distribution among tetrahedral (A) and octahedral (B) sites and the crystallite size as well. According to Neel's two-sub-lattice model, the orientation of the magnetic moment of A- and B-sites are antiparallel to each other but their spins aligned collinear, so the total magnetization is represented as $M=M_B-M_A$, where $M_A$ and $M_B$ are the magnetic moments at A- and B-sites, respectively. The $Ni^{2+}$ ions prefer to occupy B-site and $Zn^{2+}$ prefer the A-site whereas the $Fe^{3+}$ equally distributes in both A- and B-sites and $Yb^{3+}$ always prefer to go octahedral B-site. The $Fe^{3+}$ ions are replaced by the $Yb^{3+}$ substituted ions in the samples. It is seen that the $M_s$ values of substituted compositions are lower than that of parent composition (Fig. 3b). This phenomenon can be explained using change in the strength of magnetic interaction due to variation in cation distribution. To calculate the total magnetization, the following cation distribution can be acknowledged $(Zn_{0.5}\ Fe_{0.5})\ [Ni_{0.5}Fe_{2-x}Yb_x]O_4^{2-}$. The magnetic moment of $Ni^{2+}$ (2.83 $\mu_B$), $Zn^{2+}$ (0 $\mu_B$), $Fe^{3+}$ (5.96 $\mu_B$) and $Yb^{3+}$ (4.53 $\mu_B$) are calculated using the formula, $\mu_B = \sqrt{(n(n+2))}$ where *n* is the number of unpaired electron of the outermost shell of the electronic configuration. The magnetization at the B-site decreases as the concentration of $Yb^{3+}$ increases since the magnetic moment of $Yb^{3+}$ is smaller than $Fe^{3+}$. As a result, the total magnetization decreases up to *x*=0.16, then it increases [21]. The occupation of $Yb^{3+}$ on the B-site has been confirmed by the low value of calculated magnetic moment [22]. The increase of $M_s$ at *x*=0.16 may be due to creation of multi domain grains though detail is not yet understood [23].

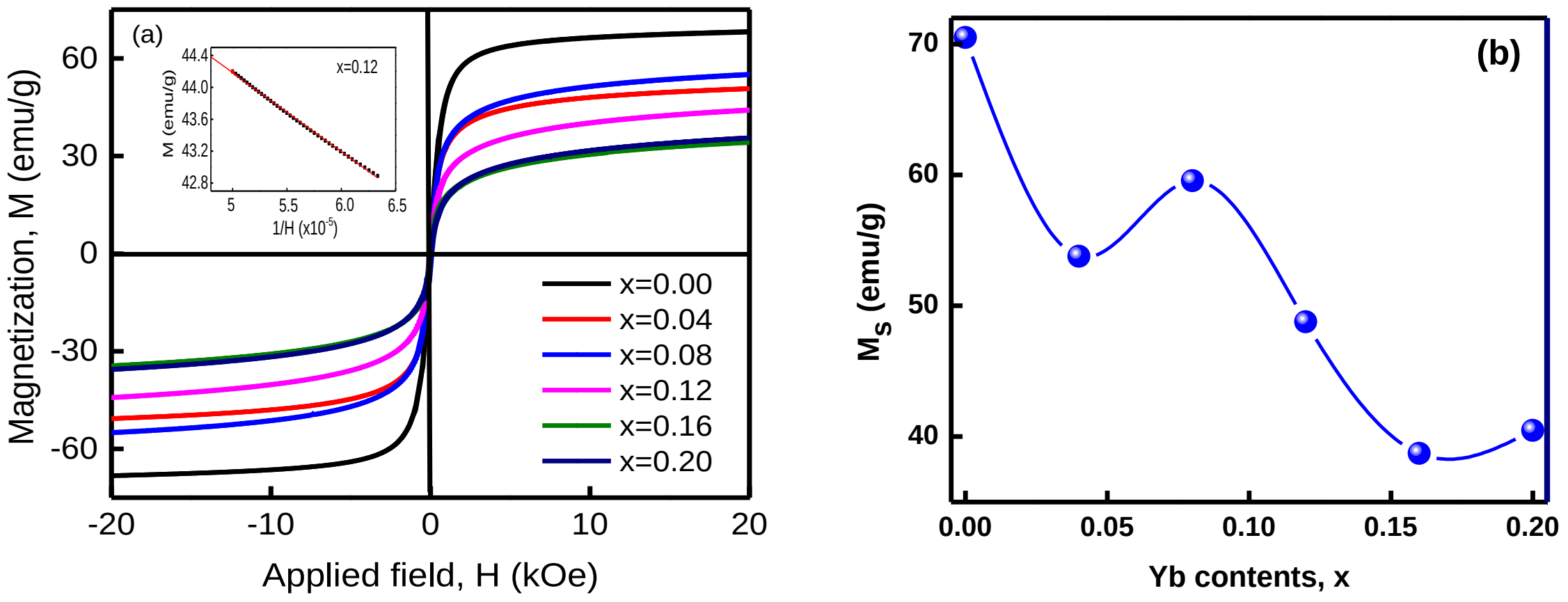

Fig 3: (a) Hysteresis (M-H) curve, the variation of (b) Yb content dependence of saturation magnetization and coercivity of $Ni_{0.5}Zn_{0.5}Yb_xFe_{2-x}O_4$ ($0 \leq x \leq 0.20$ in the step of 0.04) compositions measured at room temperature.

Table 2: Calculated magnetic parameter at room temperature of $Ni_{0.5}Zn_{0.5}Yb_xFe_{2-x}O_4$, saturation magnetization ($M_s$), coercivity (Hc), remanence magnetization (Mr), squareness ratio ($M_r/M_s$), anisotropy constant ($K_1$), calculated and theoretical magnetic moment ($\mu_B$), Y-K angle ($\alpha_{YK}$) and Curie temperature ($T_c$)

| Yb content, x | $M_s$ (emu/g) | $H_c$ (Oe) | $M_r$ (emu/g) | SQR (Mr/Ms) | $K_1$ (ergs/gm) | $\mu_B$ (cal) | $\mu_B$ (theor) | $\alpha_{YK}$ (Degree) | $T_c$ (K) |
|---|---|---|---|---|---|---|---|---|---|
| 0.00 | 70.51 | 6.77 | 0.44 | 0.006 | 487.19 | 3.00 | 7.33 | 54.62 | 550 |
| 0.04 | 53.79 | 23.09 | 1.06 | 0.020 | 1267.349 | 2.33 | 7.28 | 58.87 | 515 |
| 0.08 | 59.56 | 30.67 | 1.63 | 0.027 | 1863.92 | 2.63 | 7.22 | 56.70 | 510 |
| 0.12 | 48.77 | 47.30 | 3.30 | 0.068 | 2353.80 | 2.20 | 7.17 | 59.4 | 480 |
| 0.16 | 38.71 | 34.41 | 1.77 | 0.046 | 1359.11 | 1.78 | 7.11 | 61.97 | 415 |
| 0.20 | 40.48 | 24.17 | 1.64 | 0.041 | 998.39 | 1.89 | 7.06 | 61.05 | 450 |

Moreover, the surface to volume ratio increases due to decrease in crystallite size and the surface effect is prominent in the samples. Due to the distorted surface structure, the surface atoms are affected by strain and create vacancies. The exchange bonds on the surface are broken due to low coordination number and various interatomic spacing that induces spin disorder in the spin texture of the samples. The spin disorder reveals the low magnetization, therefore the $M_s$ decreases eventually [24, 25]. To reveal the spin disorder and domain rotation mechanism in the samples, it is essential to study the crystallite size dependence of $M_s$ and $H_c$ curves, as shown in Fig 4(a). It is seen that the $M_s$ increases with the increase in particle size or $Yb^{3+}$ ions substitution, which shows highest $H_c$ value (47.3 Oe) at $D_{xrd}$ = 6.4 nm ($x$=0.12), indicating critical particle size (or simple critical size) of the samples. That is the transition point where lower side of this value the composition exhibits single domain region and beyond the material is transferred into multi domain region [26]. The critical particle size and clear single and multi-domain rotation region have been identified for the ferrites composition of $Ni_{0.5}Zn_{0.5}Yb_xFe_{2-x}O4$ ($x$=0.00, 0.04, 0.08, 0.12, 0.16 and 0.2) sintered at 700°C in this study [Fig. 4(a)].

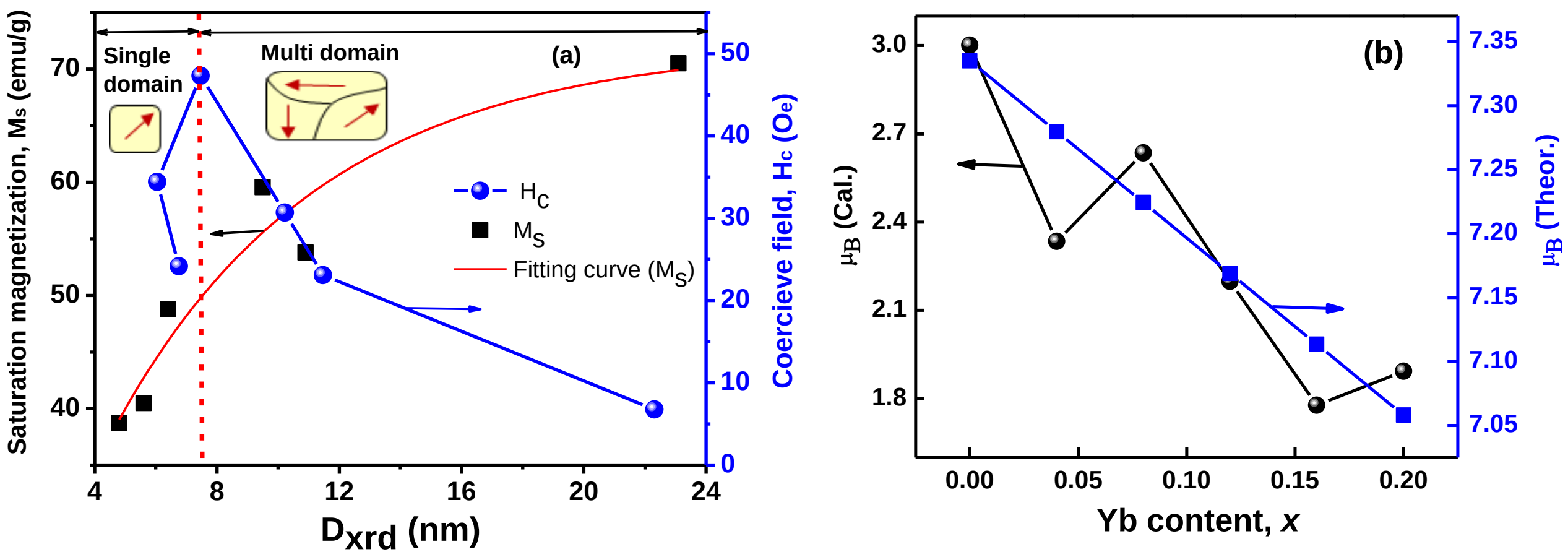

Fig 4. (a) Crystallite size dependence of saturation magnetization ($M_s$) and (b) Bohr magnetic moment as a function of $Yb^{3+}$ contents in $Ni_{0.5}Zn_{0.5}Yb_xFe_{2-x}O_4$ ($0 \leq x \leq 0.20$ in the step of 0.04) ferrites sintered at 700°C.

The magnetization or demagnetization of a magnetic material occurs due to the domain rotation. The number of grain boundaries increases due to decrease in grain size; consequently, more energy is required to magnetize or demagnetize the domain i.e., coercivity of the materials increases. It means that the coercivity is inversely related to the grain size of the materials. The grain size dependent coercivity is also calculated and illustrated in Fig. 4(a). The coercivity increases with increasing gran size as expected, except for $x$= 0 and 0.04 (24.1 and 34.4 nm), where the critical size effect is affected [27]. The smaller value of $H_c$ suggests the compositions are ferromagnetic.

The magnetic moment per formula unit ($\mu_B$) for the compositions has been calculated using the equation: $\mu_B = \frac{M \times M_s}{5585}$, where M is the molecular weight and $M_s$ is the saturation magnetization. The calculated and theoretical $\mu_B$ as a function of $Yb^{3+}$ content are shown in Fig. 4 (b). From the Fig, it is clear that calculated magnetic moment follows decreasing trend with an increasing value of $x$ similar to $M_s$ of the compositions except for $x$=0.08 where it increases. The theoretical $\mu_B$ follows typical decreasing trend with the $x$ content. In contrast, the calculated $\mu_B$ values are lying at average position and the theoretical $\mu_B$ values (Fig. 4b) are acceptable for the experimental aspect. In theoretical case, an ideal cell has been considered for evaluation. The decreasing trend of calculated Bohr magnetic moment can be explained by Neel's collinear two sub-lattice model. This nonlinear spin arrangement can also be explained by Yafet-Kittel (Y-K) three sub-lattice model. Yafet and Kittel divide the B lattice into two sub-lattices $B_1$ and $B_2$ with equal magnetic moment but oppositely canted keeping the same angle $\alpha_{YK}$ [28]. The Y-K angle provides ~~the~~ information regarding the presence of canted spin and the change of magnetic moment due to substitution $Yb^{3+}$ ions in the composition. The Y-K angle is calculated by the equation follows [29],

$$\alpha_{YK} = Cos^{-1}\frac{\mu_B + M_B}{M_A}$$

where $\mu_B$ is the calculated magnetic moment in Bohr magneton and $M_A$ and $M_B$ are the magnetic moments on the tetrahedral and octahedral sites, respectively. The calculated Y-K angles are listed in Table 2. The Y-K non-zero value suggests that the magnetic behavior could not be explained using the Neel two sub-lattice model. The configuration of spin texture on the B-sites is non-collinear type and the interaction between A- and B-sites decreases with increasing substitution ions as well [27]. This phenomenon is also responsible for the decrease of $M_s$ with increasing content of $Yb^{3+}$.

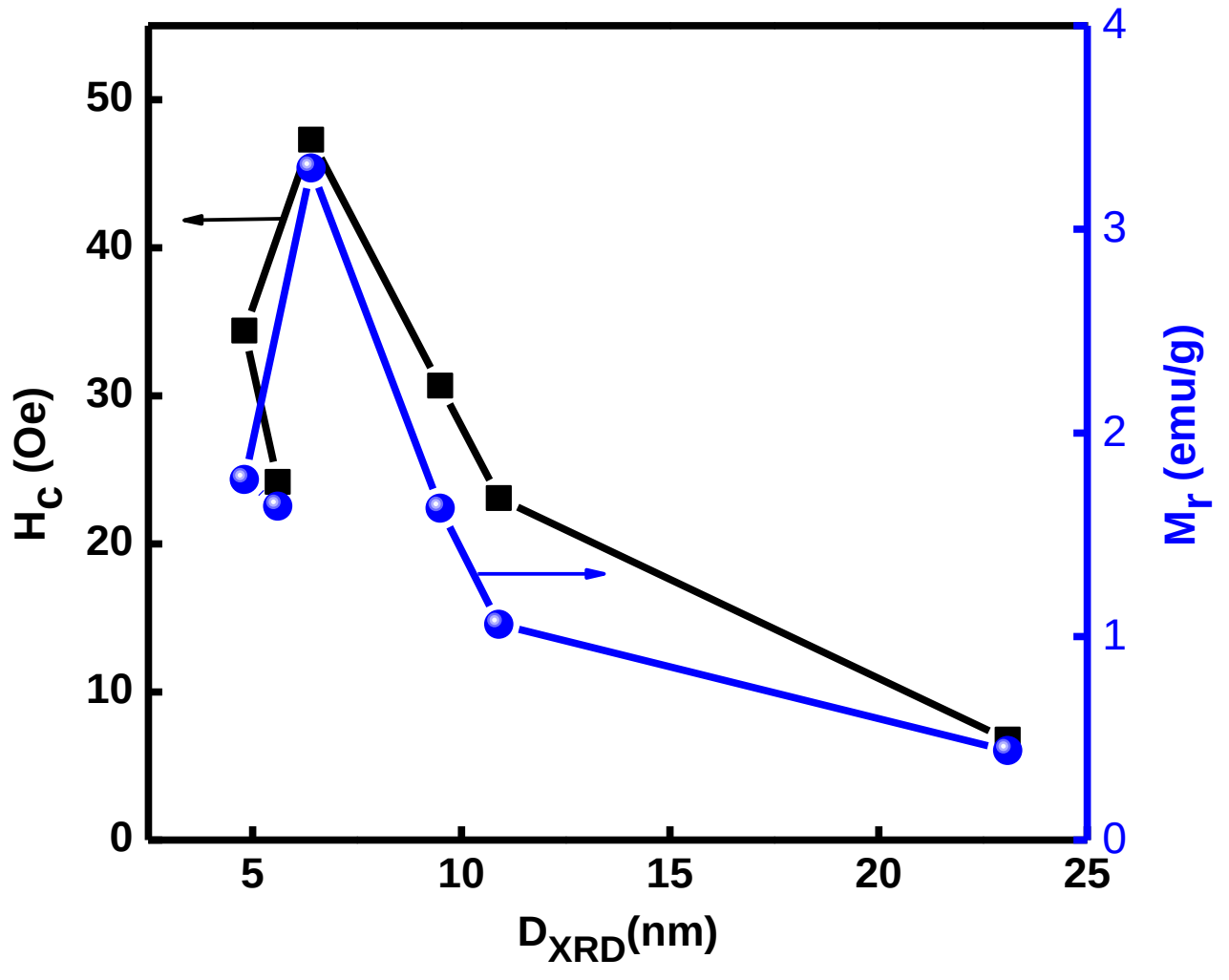

Fig. 5: Crystallite size dependence of coerceive field ($H_c$) and remanent magnetization ($M_r$) as a function of $Yb^{3+}$ contents in the $Ni_{0.5}Zn_{0.5}Yb_xFe_{2-x}O_4$ ($0 \leq x \leq 0.20$ in the step of 0.04) compositions sintered at 700°C.

The characteristic parameter remanent magnetization ($M_r$) of the composition and the coercivity field ($H_c$) as a function of average grain size have been depicted in Fig. 5. Both the parameters show similar trend, i.e., initially increases (due to strong spin interaction during spin alignment of the nanoparticles [30]) and after critical size at $D_{xrd}$ = 6.4 nm ($x$=0.08), it decreases with crystallite size and crossover from single domain regime to multi domain regime. A single domain crystal spontaneously breaks into number of domains. This reduces the magnetization energy and the crystallite prefers to remain a single domain in a quite small particle size. In the single domain regime, the dependence of coercivity with the particle size can be written as,

$H_c = g - \frac{h}{D^{3/2}}$ where $g$, h and D, $H_c$ are the constants and the particle size, coercivity, respectively [25]. The equation shows that the $H_c$ increases with increasing crystallite size below a critical size. In the multi-domain regime, this equation modifies based on the experimental observation as, $H_c = a + \frac{b}{D}$, where $a$ and b are constants. Therefore, beyond the critical size the coercivity decreases with increasing the crystallite size [25].

The squareness ratio ($SQR = M_r/M_s$) provides the information regarding whether nanoparticles are multiple magnetic domain (MMD) [SQR value is <0.5] [31, 33] or single magnetic domain (SMD) [SQR value is <0.5] [2, 33, 34]. The materials with a higher SQR value make them plausible candidates for applying magnetic recording and memory devices [35]. In Table 2, the SQR values decrease with increasing $Yb^{3+}$ ions indicating the surface spin disordering is very strong in the compositions under study. Moreover, the SQR values are found to be in the range of 0.006 to 0.068, suggesting the compositions are in the MMD structure. The very small SQR value demonstrates that some ~~of~~ nanoparticles are relaxing so fast such as super paramagnet at RT, even in the absence of an external magnetic field. Zero value of $H_c$ has not been observed in Fig. 4(a) consequently, superparamagnetic effect could not be claimed. According to SQR values, all compositions are in the MMD regime, this is not contradictory with previous paragraph where $H_c$ characterized the compositions are divided into SSD and MMD regime. This might be due to superparamagnetic effect which actives below the estimated critical size (6.4 nm) therefore, the SQR does not show higher value.

The magnitude of the anisotropy constants $K_1$ is defined as the strength of the anisotropy in any particular crystal. The anisotropy constants are calculated using the following equation [35], $H_c = \frac{(0.98 \times K1)}{Ms}$, here $H_c$ and $M_s$ denote the coercivity and saturation magnetization and displayed in Table 2. It is seen that the value of $K_1$ increases up to $x$= 0.12 (487 to 2353) and after that it decreases due to the effect of $H_c$ since the anisotropy strongly contributes to the $H_c$ [25]. The spin-orbital coupling may also influence the anisotropy at the atomic level of the $Yb^{3+}$.

*Correlation between bond length, bond angle and magnetic properties*

The bond length (cation-anion bond: p, q, r and s and cation-cation bond: b, c, d, e and f) and bond angles ($\theta_1$, $\theta_2$, $\theta_3$, $\theta_4$ and $\theta_5$) play a vital role in the strength of different exchange interactions among the ions of the composition. The angles $\theta_1$ and $\theta_2$, $\theta_3$ and $\theta_4$, and $\theta_5$ are responsible for the A-O-B, B-O-B and A-O-A exchange interaction, respectively. The bond lengths and angles are calculated using the following formula [36] and shown in Fig. 6.

$p = \left\{\left(\frac{5}{8}\right) - u\right\}$, $q = a\left(u - \frac{1}{4}\right)\sqrt{3}$, $r = a\left(u - \frac{1}{4}\right)\sqrt{11}$ and $s = a\{(1/3u) + (1/8)\}\sqrt{3}$

$$b = (a/4)\sqrt{2}\,;\ \ c = (a/8)\sqrt{11}\ ;d = (a/4)\sqrt{3}\,;e = (3a/8)\sqrt{3}\,;f = (a/4)\sqrt{6}$$

$\theta_1 = cos^{-1}[(p^2 + q^2 - c)/2pq]$, $\theta_2 = cos^{-1}[(p^2 + r^2 - e^2)/2pr]$, $\theta_3 = cos^{-1}[(2p^2 - b^2)/2p^2]$, $\theta_4 = cos^{-1}[(p^2 + s^2 - f^2)/2ps]$ and $\theta_5 = cos^{-1}[(r^2 + q^2 - d^2)/2rq]$

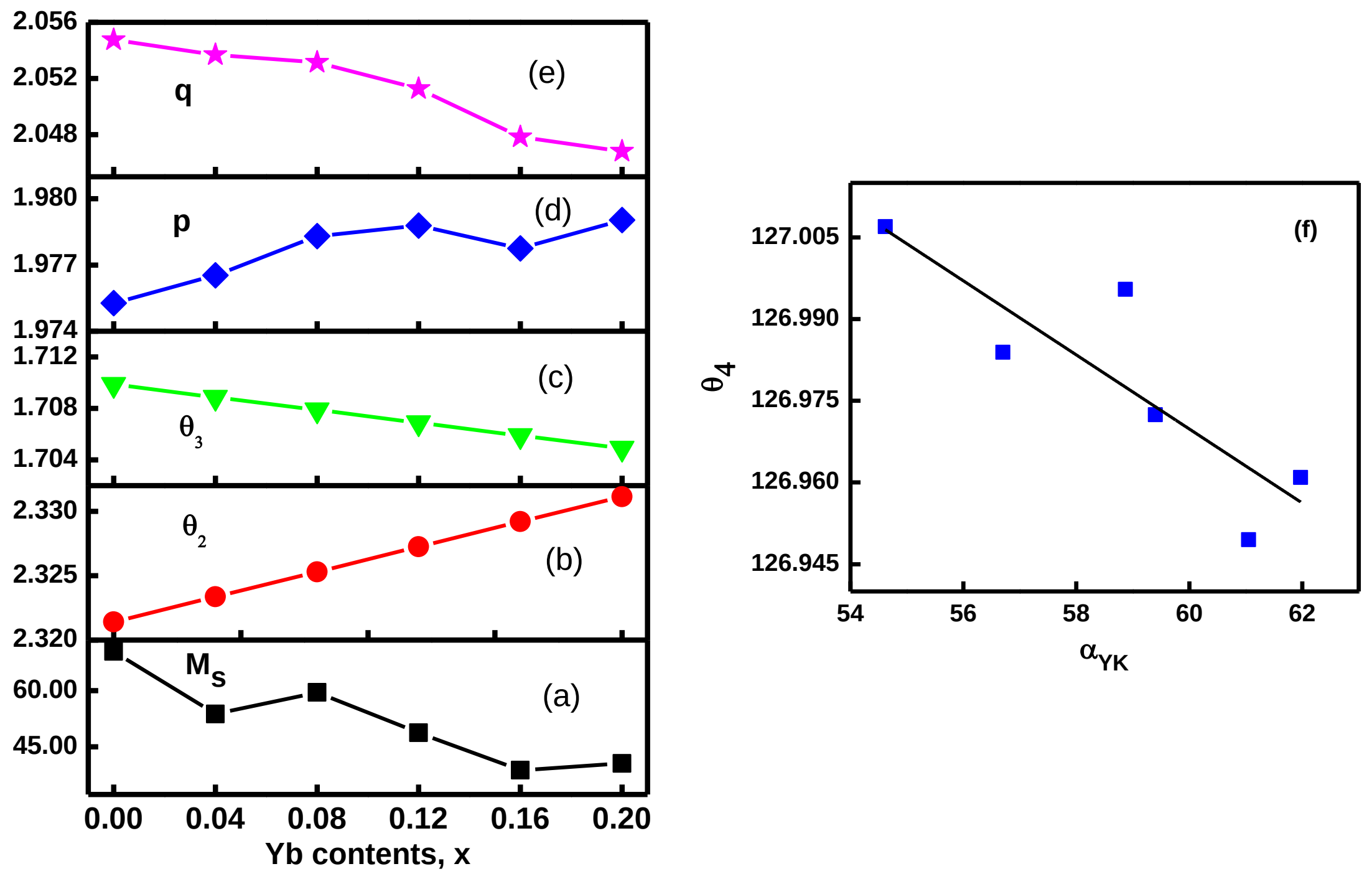

Fig. 6. Correlation between Yb content and (a) saturation magnetization ($M_s$), (b) bond angle (A-O-B) $\theta_2$, (c) bond angle (B-O-B) $\theta_3$, cation-anion distance (d) at octahedral site, p and (e) tetrahedral site, q. (f) variation of bond angle $\theta_4$ with canted angle $\alpha_{YK}$ (solid line is the linear fit) as a function of substitution concentration of $Yb^{3+}$ in $Ni_{0.5}Zn_{0.5}Yb_xFe_{2-x}O_4$ ($0 \leq x \leq 0.20$ in the step of 0.04) compositions sintered at 700°C.

The strengthening of A-O-B exchange interaction occurs since the angle $\theta_1$, $\theta_2$ increases and the cation-anion distance at tetrahedral site (q) decreases with $Yb^{3+}$ contents. However, the bond angle $\theta_3$, $\theta_4$ decreases and the bond length between cation-anion at octahedral site (p) increases, indicating weakening of B-O-B interaction [37]. Fig. 6(f) illustrated the variation of $\theta_4$ as a function of the canted angle $\alpha_{YK}$. It shows that B-O-B interaction (due to bond angle $\theta_4$) decreases with increasing the canted angle along with increases the A-O-B exchange interaction in the composition consequently the net $M_s$ are varied with $Yb^{3+}$ content [29].

*Curie temperature*

In general temperature significantly affects the magnetic moment and rapid decline of magnetic moment in close vicinity to the transition temperature from the ferrimagnetic (FM) to paramagnetic (PM) state occurs. Consequently, the transition temperature (or Curie temperature, $T_c$) can be determined using the magnetic moment vs. temperature (M-T) graph. To determine the $T_c$, magnetic field (100 Oe) was applied to the samples and resultant magnetic moments are recorded as a function of temperature. Low heating rate was maintained to avoid growing access particles during heating the sample. The M-T graph has been measured in the temperature range of 300 to 700 K and is shown in Fig. 7 (a). The magnetic moment decreases with increasing temperature as expected and the magnetic moment tends to zero at a certain temperature for a specific composition. The first derivative of magnetic moment against temperature (dM/dT vs. T in Fig. 7b) graph is used to determine the $T_c$ value and depicted in Fig. 7 (c). The lowest of dM/dT vs T provides the values of $T_c$ [Fig. 7c] where the transition from the FM to PM has taken place. Fig. 7(c) shows that the $T_c$ follows the decreasing trend with increasing the concentration of $Yb^{3+}$ except $x$=0.20 where it is bit increased however still smaller than undoped ($x$=0) Ni-Zn ferrite. It is well known that substitutional disorder always causes local strains and inhomogeneities inside the compounds, which lead to a distribution in the exchange energies. According to this mean-field approximation, the $T_c$ is directly proportional to the exchange energy [38].

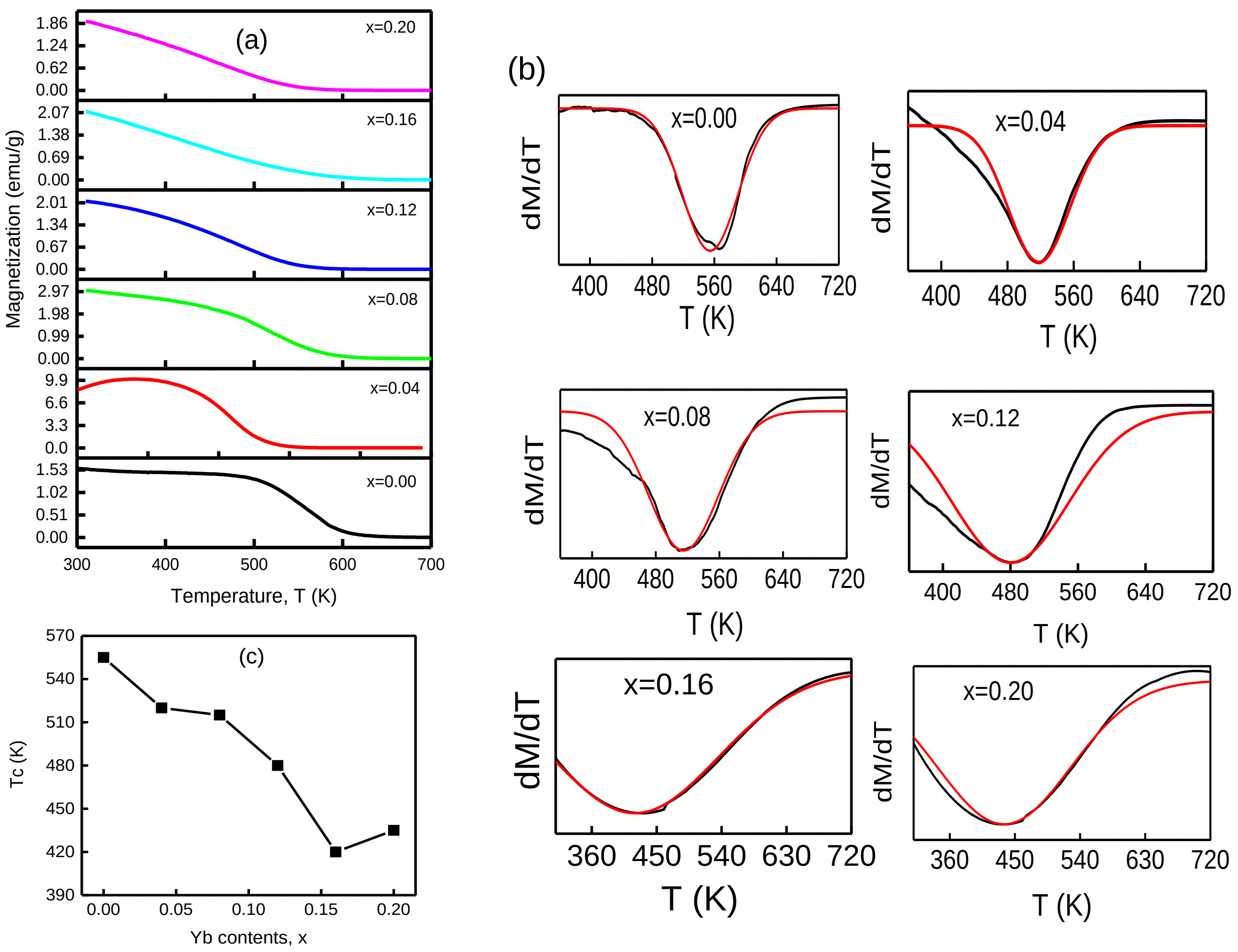

Fig. 7: (a) Temperature dependent magnetization (M-T) curve, (b) temperature dependent dM/dT curves for different Yb contents and (c) variation of Curie temperature with Yb concentration in the $Ni_{0.5}Zn_{0.5}Yb_xFe_{2-x}O_4$ compositions sintered at 700°C.

*Permeability*

Real part of permeability ($\mu_i'$) represents the ability of the production of magnetic field within the materials or due to applied magnetizing field the amount of relative increment or decrement of the induced magnetic field within the materials. However, the imaginary part ($\mu_i''$ ) represents the energy lost in the materials and shows the out of phase relationship of B with H. The frequency dependence of complex permeability (a) real part ($\mu_i'$) and (b) imaginary part $(\mu_i'')$, (c) relative quality factor (RQF) and (d) the values of $\mu_i'$ at ~ 20 Hz frequency for different $Yb^{3+}$ contents of compositions $Ni_{0.5}Zn_{0.5}Yb_xFe_{2-x}O_4$ ($x$=0.00, 0.04, 0.08, 0.12, 0.16 and 0.2) sintered at 700 °C are presented in Fig. 8, respectively.

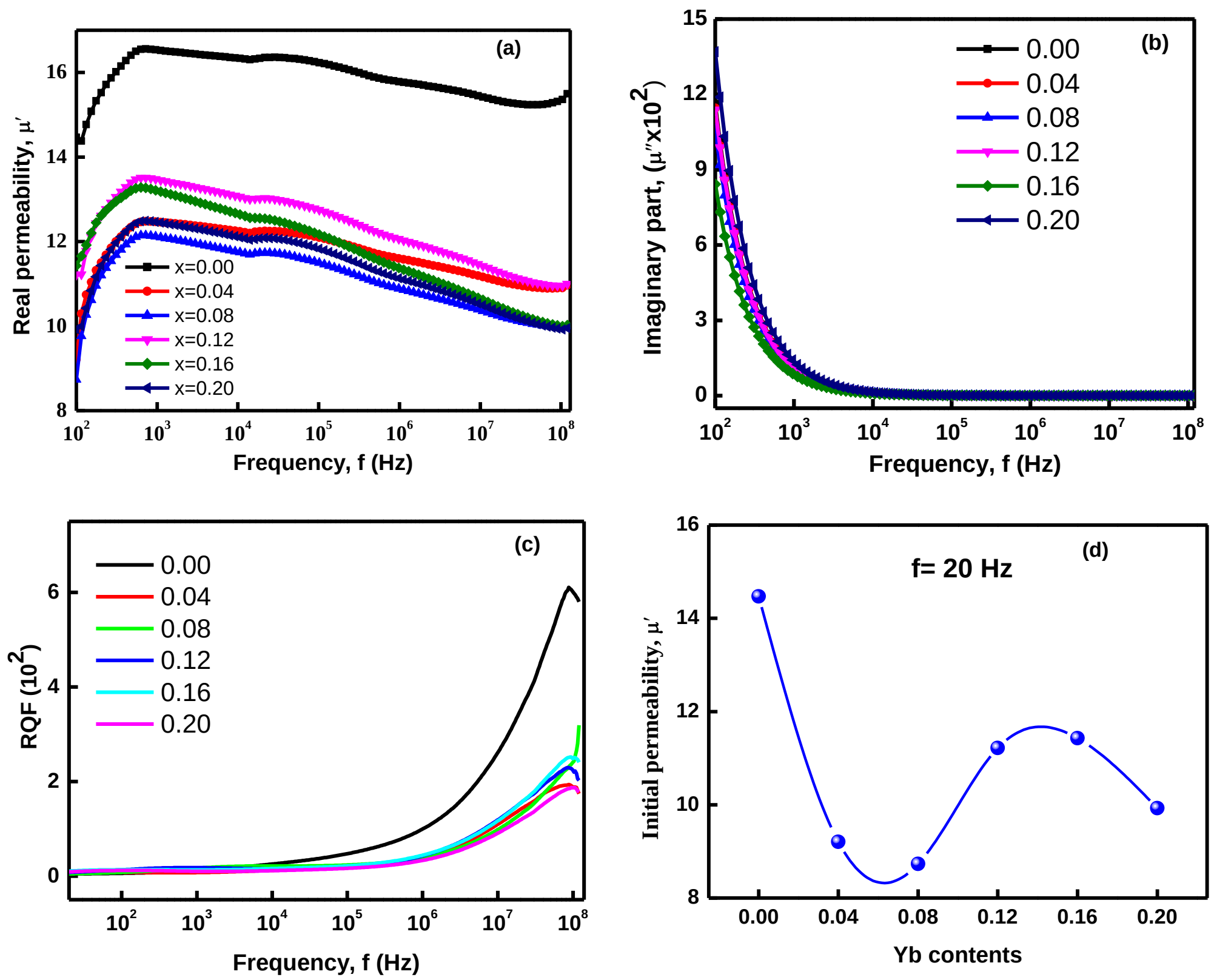

Fig. 8: Frequency dependence of (a) real part and (b) imaginary part of permeability of Yb substituted Ni-Zn ferrites (c) relative quality factor (RQF) and (d) Yb content dependent of $\mu_i'$ at ~ 20 Hz.

It is seen for all samples that $\mu_i'$ increases up to certain frequency (~600 Hz) to reach a maximum and then it decreases up to more than $10^8$ Hz [Fig. 8 (a)]. Two established mechanisms such as domain wall motion and spin rotation occur in the phenomena of permeability. The contribution of domain wall motion reduces at lower frequency region since the domain wall motion depends on the square of the frequency consequently the $\mu_i'$ rises. However, the spin rotation is inversely proportional to the frequency therefore the value of $\mu_i'$ declines at higher frequency [37]. The value of imaginary part $\mu_i''$ rapidly declines with increasing applied frequency and after certain a frequency, it is almost constant at around zero value. The ferrimagnetic resonance is not found at a particular frequency where a prominent peak is observed at the frequency where the $\mu_i'$ starts to decrease [39]. We could not observe such a type of peak. However, we repeat these measurements several times and reproduce the results and reasons that have not been uncovered yet and the study is under consideration.

The relative quality factor (RQF) is an essential factor for measuring the performance of the samples from the application point of view. Frequency dependent RQFs for the titled compositions are calculated using the relation, $RQF = \frac{\mu_i'}{tan\delta}$ and are illustrated in Fig. 8(c). A very low value of *RQF* has been observed at low frequency and found to be increased with frequency and reach the maximum value to show a characteristics peak [40] then it should be decreased to low value at higher frequency. It seems that a complete peak would appear after ~$10^8$ MHz frequency range that is out of our instrument range therefore we could not see the range even the peak as well. The maximum value of RQF is found to be $x$ = 0.0 and peak at very high frequency for x=0.08 compositions since the RQF is strongly dependent on the microstructure (i.e., grain size, porosity, etc.) (Table 1). The values of $\mu_i'$ at ~ 20 Hz frequency for different $Yb^{3+}$ contents are depicted in Fig. 8 (d). The dependency of $\mu_i'$ on Yb content can be explained by the following relation [41]: $\mu_i' \propto \frac{M_S^2 D}{\sqrt{K_1}}$ where $\mu_i$ is the initial permeability, $M_s$ is the saturation magnetization, $D$ is the average grain size and $K_1$ is the magneto-crystalline anisotropy constant. The value of $\mu_i'$ is proportional to the square of $M_s$ (Fig. 3b) therefore, observed trend of $\mu_i'$ with Yb contents (Fig. 8c) is expected.

M*agnetic properties mapping at high temperature*

The materials that retain and exhibit magnetic ordering at high temperature are essential properties to be applied in high temperature magnetic devices. To evaluate the feasibility of the compositions, the magnetization vs. the applied magnetic field (upper quadrant of hysteresis loops) at various measuring temperatures are measured and illustrated in Fig. 9 (a-f). It is seen that the magnetic order is smashed above $T_c$, i.e., transition from ferrimagnetic (below $T_c$) to paramagnetic (beyond $T_c$) regime. It is noteworthy that the compositions remain the ferrimagnetic ordered structured up to $T_c$ indicating these materials are plausible candidates to use at higher temperature magnetic devices.

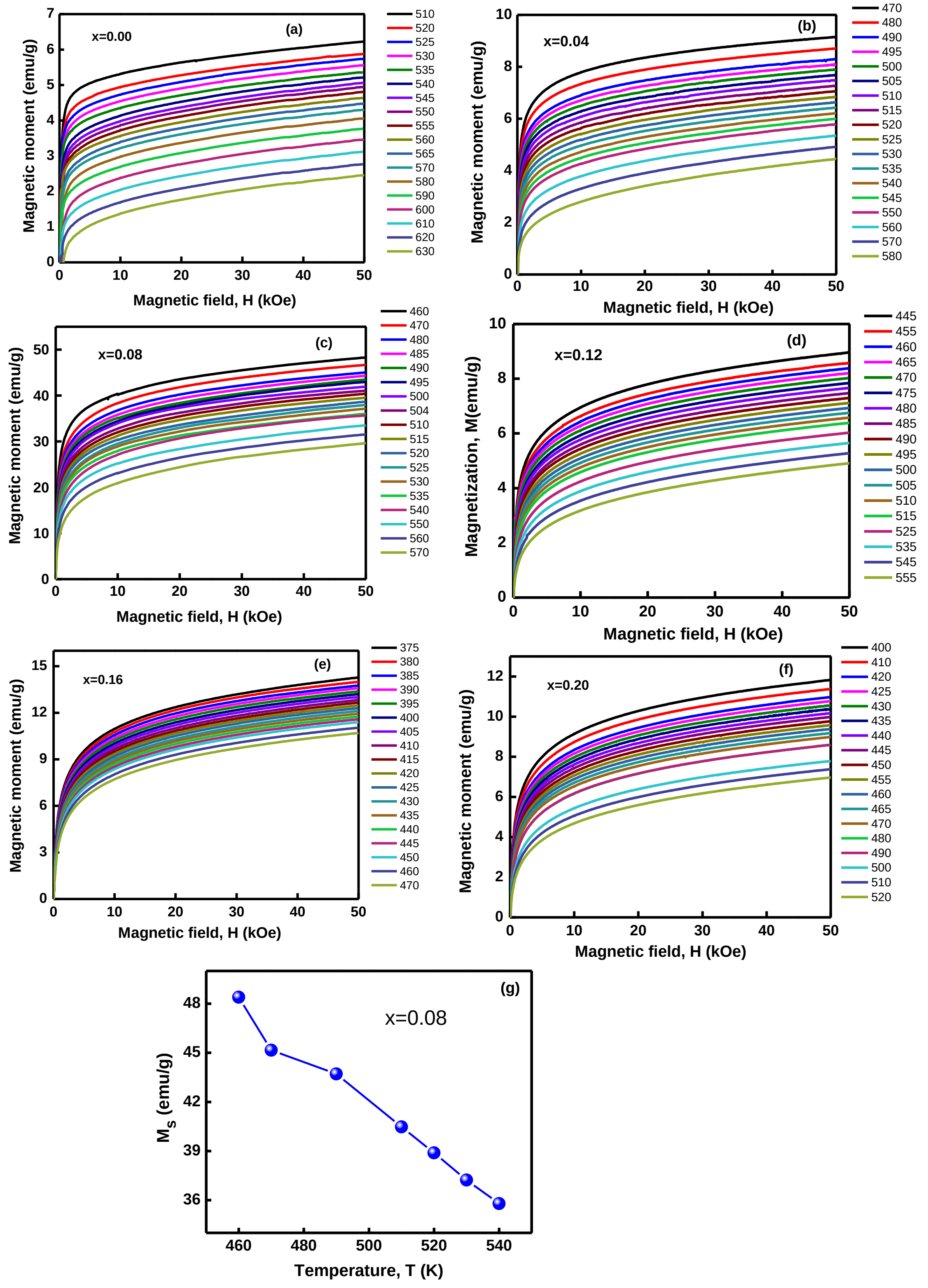

Fig. 9: (a-f) upper quadrant of M-H curve of $Ni_{0.5}Zn_{0.5}Yb_xFe_{2-x}O_4$ ferrites at different measuring temperature, (g) variation of $M_s$ as a function of measuring temperature for $x$=0.08.

The value of $M_s$ for each composition at selected temperatures has been measured as $M_s$ vs. 1/H curves are plotted (figures not shown here) and are extrapolated to zero, i.e. 1/H = 0 [20]. The $M_s$ values are following decreasing trend with increasing measuring temperature for all compositions. A typical plot of $x$=0.08 is illustrated in Fig. 9 (g). The randomness of magnetic moment and change in site occupancy of cation is responsible for decreasing the $M_s$ with increasing temperature ascribed from the increase in thermal energy and their fluctuation at higher temperatures.

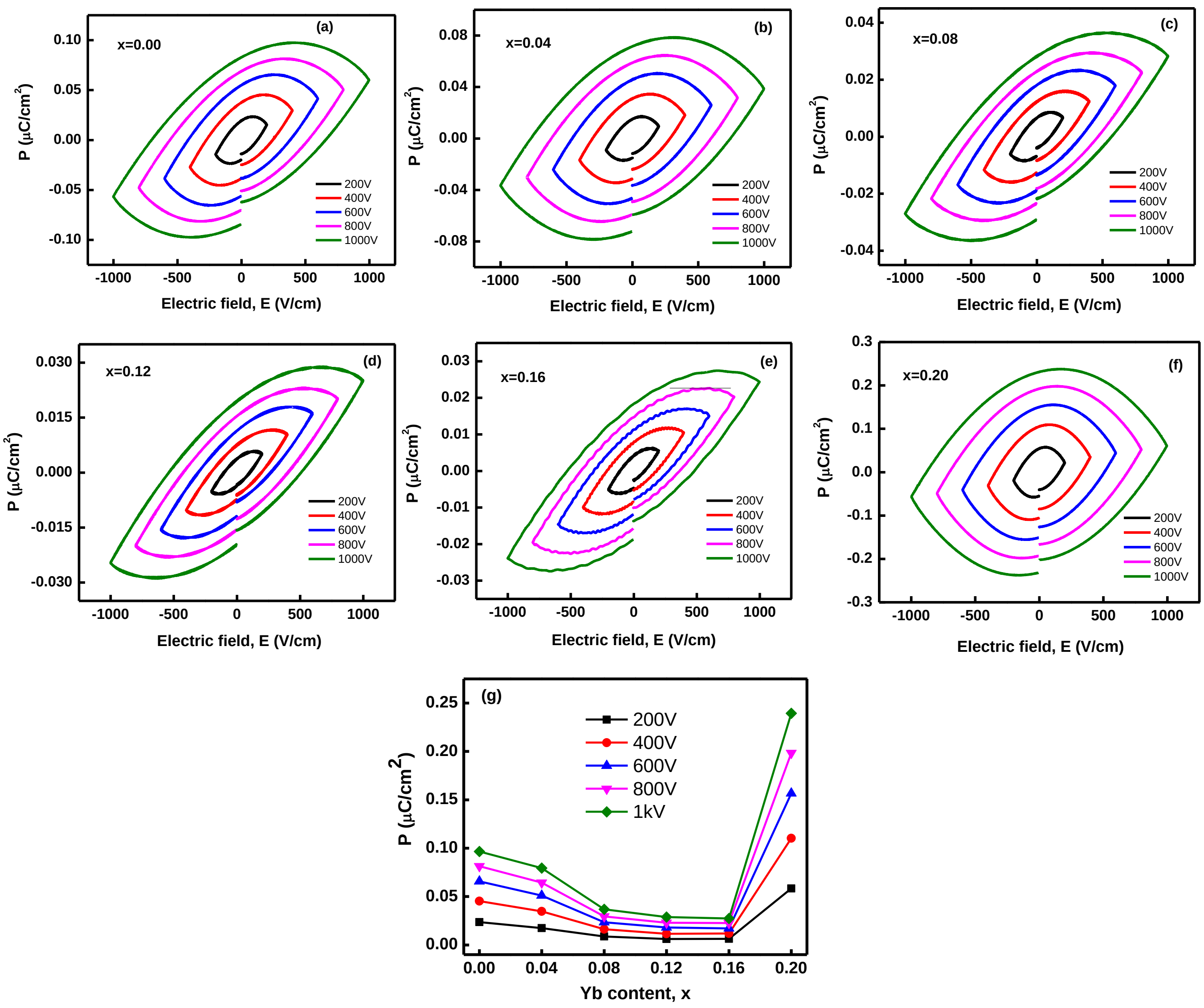

Fig. 10: (a-f) The ferroelectric loop (P-E) for different compositions at various electric field (g) Yb content dependent maximum polarization at various applied voltages.

The ferroelectric hysteresis loop (Polarization-Electric field, P-E) of all the samples at selective applied electric fields 200, 400, 600, 800 and 1000V $cm^{-1}$ are measured at RT and represented in Fig. 10. The P-E loop confirms the ferroelectric nature of the studied samples. It is seen that the hysteresis loops are unsaturated which arise due to the leakage current in the composition. This leakage current might ascend from the presence of oxygen vacancies in the materials. The compositions follow the trend that the remanent polarization and coercivity rise with increasing applied electric field from 200 V to 1 kV. However, the ferroelectric parameter (P) decreases with the concentration of the substituent ($Yb^{3+}$) upsurges except for $x$=0.20 where it is increased abruptly. This decline is due to the compact reduced phase fraction of ferroelectric phase since the ferroelectric domain rearranges in the direction of applied electric field, consequently diminishing the ferroelectric properties. The ferroelectric properties (standard P-E loop) of the titled compositions make them a credible candidate for model electronic devices [42].

## 4. Conclusions

Yb-substituted $Ni_{0.5}Zn_{0.5}Yb_xFe_{2-x}O_4$ ($x$=0.00, 0.04, 0.08, 0.12, 0.16 and 0.2) ferrites sintered at 700 °C have been successfully synthesized using sol-gel auto combustion route. Single phase cubic spinel structure has been confirmed by the XRD pattern. The crystallite size and average grain size decreases from 23 to 5 nm and 52 to 17 nm, respectively. Saturation magnetization ($M_s$) decreases from 70 to 38 emu/g and coercivity increases 6 to 24 Oe with the increase in $Yb^{3+}$ contents. The $M_s$ increases with increasing particle size/$Yb^{3+}$ ions, showing the maximum value of $H_c$ is of 47.3 Oe at $D_{xrd}$ 6.4 nm indicating the critical particle size of the compositions which is the transition point of single domain regime and beyond it is transferred into multi domain regime. The very small values of squareness ratio, SQR (0.006 to 0.068) demonstrates that some nanoparticles are relaxing so fast such as super paramagnet at room temperature even in the absence of an external magnetic field. The value of $K_1$ increases from 487 to 2353 up to $x$= 0.12, after that it decreases due to the effect of $H_c$. Strengthening of A-O-B and weakening of B-O-B exchange interaction occur in the studied compositions and $T_c$ decreases from 550 to 415 K. The variation of $\mu_i'$ has been observed for all compositions and successfully explained by the established domain wall and spin rotation mechanism. It is noteworthy that the composition remains the ferrimagnetic ordered structured up to $T_c$, indicating these materials are a plausible candidate to use at high temperature devices. High frequency peak of relative quality factor ($10^8$ Hz) indicating the compositions are suitable for high frequency applications. Standard Polarization-Electric field, P-E loops for all samples have been observed.

## Acknowledgments

The authors are grateful to the Directorate of Research and Extension (DRE), Chittagong University of Engineering and Technology (CUET), Chattogram-4349, Bangladesh, for arranging financial assistance under grant number CUET DRE (CUET/DRE/2016-2017/PHY/003) for arranging the financial support for this work. We are also thankful for the laboratory support of the Materials Science Division, Atomic Energy Centre, Dhaka 1000, Bangladesh.